\documentclass[
twocolumn,prd,
showpacs,
nofootinbib,
amsmath,amssymb,
superscriptaddress]{revtex4-1}

\usepackage{float}
\usepackage{graphicx}% Include figure files                                                                                                                                                           
\usepackage{dcolumn}% Align table columns on decimal point                                                                                                                                            
\usepackage{bm}% bold math                                                                                                                                                                            
\usepackage{hyperref}% add hypertext capabilities                                                                                                                                                     
\usepackage{ulem} %xout command                                                                                                                                                                       
\usepackage{gensymb}
\usepackage{subcaption}

\newcommand{\be}{\begin{equation}}
\newcommand{\ee}{\end{equation}}
\newcommand{\bi}{\begin{itemize}}
\newcommand{\ei}{\end{itemize}}
\newcommand{\bea}{\begin{eqnarray}}
\newcommand{\eea}{\end{eqnarray}}

\usepackage{color}

\usepackage{soul}

\begin{document}

\title{Mayawaves: Python Library for Interacting with the Einstein Toolkit and the MAYA Catalog}

\affiliation{Center for Gravitational Physics and Department of Physics, The University of Texas at Austin, Austin, TX 78712}

\author{Deborah Ferguson$^{1}$}\noaffiliation
\author{Surendra Anne$^{1}$}\noaffiliation
\author{Miguel Gracia-Linares$^{1}$}\noaffiliation
\author{Hector Iglesias$^{1}$}\noaffiliation
\author{Aasim Jan$^{1}$}\noaffiliation
\author{Erick Martinez$^{1}$}\noaffiliation
\author{Lu Lu$^{1}$}\noaffiliation
\author{Filippo Meoni$^{1}$}\noaffiliation
\author{Ryan Nowicki$^{1}$}\noaffiliation
\author{Max L. Trostel$^{1}$}\noaffiliation
\author{Bing-Jyun Tsao$^{1}$}\noaffiliation
\author{Finny Valorz$^{1}$}\noaffiliation

{
\let\clearpage\relax
\maketitle
}

\section*{Summary}
Einstein’s Theory of General Relativity (GR) dictates how matter responds to the cur- vature of space-time and how space-time curves due to matter. From GR came the prediction that orbiting massive objects would create ripples in space-time called gravi- tational waves (GW). In 2015, the Laser Interferometer Gravitational-Wave Observatory (LIGO)~\cite{LIGOScientific:2014pky} detected the first such GW signal from merging binary black holes (BBHs)~\cite{LIGOScientific:2016aoc}, and in the years since, the LIGO, Virgo, and KAGRA Collaborations (LVK) have accumulated 90 detections of merging compact objects~\cite{VIRGO:2014yos, KAGRA:2020cvd, LIGOScientific:2018mvr, LIGOScientific:2020ibl, LIGOScientific:2021djp}.  Extracting these signals from noise and using them to infer the parameters of coalescing black holes (BHs) relies upon having vast template banks that accurately predict the expected GWs~\cite{LIGOScientific:2017vwq, LIGOScientific:2020aai, Shibata:2017xdx, Hannam:2013oca, Bohe:2016gbl, Khan:2015jqa, Blackman:2017pcm, Husa:2015iqa, LIGOScientific:2016fbo, Lange:2017wki, Schmidt:2017btt}.

While analytic solutions exist for the simplest cases within GR, e.g. single BHs, merging BBHs have no analytic solution. Approximate methods can be used when the BHs are far apart or have highly unequal masses, but the coalescence of BHs of comparable mass must be solved computationally. Numerical relativity (NR) simulations accomplish this by evolving a BBH space-time on supercomputers, enabling us to study the dymanics of BBH systems as well as predict the GWs they emit. The Einstein Toolkit (\texttt{ETK}) is a set of tools created to perform these NR simulations~\cite{Loffler:2011ay}, and \texttt{MAYA} is a branch of \texttt{ETK} used by the MAYA collaboration~\cite{Herrmann:2006ks, Jani:2016wkt, Vaishnav:2007nm, Healy:2009zm, Pekowsky:2013ska}.

These tools allow us to study the coalescence of compact objects, their evolution, and the gravitational radiation they emit. The \texttt{Mayawaves} library introduced in this paper is an analysis pipeline used to process and analyze such NR simulations.

\section*{Statement of need}
NR simulations are crucial for studying BHs and have been instrumental in the detection of GWs by the LVK. However, these simulations produce vast amounts of data that must be processed in order to perform studies, create models, and use them with GW detection pipelines. Additionally, given the complexity of these simulations, they are typically performed for many days or weeks across many processors, leading to data which is split into several output directories and files. Sifting through all this data can be overwhelming for newcomers to the field and is cumbersome for even the most experienced numerical relativists. While it is often important to develop an understanding of these files and their complexities, in many situations, a simpler, more streamlined workflow is appropriate.

\texttt{Mayawaves} is an open-source python library for processing, studying, and exporting NR simulations performed using \texttt{ETK} and \texttt{MAYA}. When using the library to interact with a simulation, the user does not need to be familiar with all the types of output files generated by the simulation, but rather, can think in terms of physical concepts such as \textit{coalescences} and \textit{compact objects}.

The \texttt{Coalescence} class is the fundamental basis for \texttt{Mayawaves}. It represents the entirety of the BBH coalescence and serves as the interface between the user and the simulation data. The main data format used with \texttt{mayawaves} is an h5 file constructed from the raw simulation data. With this h5 file in hand, the user need only create a \texttt{Coalescence} object and then proceed with analyzing the data.

Each \texttt{Coalescence} object contains \texttt{CompactObjects} associated with each of the merging bodies as well as any remnant object. Through these \texttt{CompactObjects}, the user can track the objects’ positions, spins, masses, etc. All radiative information is stored within the \texttt{RadiationBundle} class. Each \texttt{Coalescence} object contains a \texttt{RadiationBundle} and uses it to compute gravitational wave strain, energy radiated, etc.

A number of utility modules are included to create effortless workflows that can move from raw simulations to community standard formats. A typical workflow would involve using the \texttt{PostProcessingUtils} functions to create the h5 file from the raw simulation data, using the \texttt{Coalescence} class to read that h5 file and analyze the simulation, and finally exporting the \texttt{Coalescence} object to another format such as that required by the LVK catalog~\cite{Schmidt:2017btt}.

\texttt{Mayawaves} is also the primary way to interact with the MAYA Public Catalog of NR waveforms hosted at https://cgp.ph.utexas.edu/waveform. The simulations are stored in the \texttt{Mayawave}s h5 file structure, and can be read using the \texttt{Coalescence} class. \texttt{Mayawaves} has a \texttt{CatalogUtils} module for interacting with the MAYA waveform catalog. This module includes functions for accessing and plotting the metadata for the entire catalog as well as functions to download simulations from the catalog.

\texttt{Mayawaves} is open source and is designed to be easily extensible, and we look forward to additional contributions from the ETK community.

\paragraph*{\textbf{Acknowledgements}}
The authors thank Deirdre Shoemaker and Pablo Laguna for their support throughout the duration of this code development. The work presented in this paper was possible due to grants NASA 80NSSC21K0900, NSF 2207780 and NSF 2114582.

%\bibliography{refs,nlst_wp}
\bibliography{ms}

\begin{thebibliography}{24}%
\makeatletter
\providecommand \@ifxundefined [1]{%
 \@ifx{#1\undefined}
}%
\providecommand \@ifnum [1]{%
 \ifnum #1\expandafter \@firstoftwo
 \else \expandafter \@secondoftwo
 \fi
}%
\providecommand \@ifx [1]{%
 \ifx #1\expandafter \@firstoftwo
 \else \expandafter \@secondoftwo
 \fi
}%
\providecommand \natexlab [1]{#1}%
\providecommand \enquote  [1]{``#1''}%
\providecommand \bibnamefont  [1]{#1}%
\providecommand \bibfnamefont [1]{#1}%
\providecommand \citenamefont [1]{#1}%
\providecommand \href@noop [0]{\@secondoftwo}%
\providecommand \href [0]{\begingroup \@sanitize@url \@href}%
\providecommand \@href[1]{\@@startlink{#1}\@@href}%
\providecommand \@@href[1]{\endgroup#1\@@endlink}%
\providecommand \@sanitize@url [0]{\catcode `\\12\catcode `\$12\catcode
  `\&12\catcode `\#12\catcode `\^12\catcode `\_12\catcode `\%12\relax}%
\providecommand \@@startlink[1]{}%
\providecommand \@@endlink[0]{}%
\providecommand \url  [0]{\begingroup\@sanitize@url \@url }%
\providecommand \@url [1]{\endgroup\@href {#1}{\urlprefix }}%
\providecommand \urlprefix  [0]{URL }%
\providecommand \Eprint [0]{\href }%
\providecommand \doibase [0]{http://dx.doi.org/}%
\providecommand \selectlanguage [0]{\@gobble}%
\providecommand \bibinfo  [0]{\@secondoftwo}%
\providecommand \bibfield  [0]{\@secondoftwo}%
\providecommand \translation [1]{[#1]}%
\providecommand \BibitemOpen [0]{}%
\providecommand \bibitemStop [0]{}%
\providecommand \bibitemNoStop [0]{.\EOS\space}%
\providecommand \EOS [0]{\spacefactor3000\relax}%
\providecommand \BibitemShut  [1]{\csname bibitem#1\endcsname}%
\let\auto@bib@innerbib\@empty
%</preamble>
\bibitem [{\citenamefont {Aasi}\ \emph {et~al.}(2015)\citenamefont {Aasi} \emph
  {et~al.}}]{LIGOScientific:2014pky}%
  \BibitemOpen
  \bibfield  {author} {\bibinfo {author} {\bibfnamefont {J.}~\bibnamefont
  {Aasi}} \emph {et~al.} (\bibinfo {collaboration} {LIGO Scientific}),\ }\href
  {\doibase 10.1088/0264-9381/32/7/074001} {\bibfield  {journal} {\bibinfo
  {journal} {Class. Quant. Grav.}\ }\textbf {\bibinfo {volume} {32}},\ \bibinfo
  {pages} {074001} (\bibinfo {year} {2015})},\ \Eprint
  {http://arxiv.org/abs/1411.4547} {arXiv:1411.4547 [gr-qc]} \BibitemShut
  {NoStop}%
\bibitem [{\citenamefont {Abbott}\ \emph
  {et~al.}(2016{\natexlab{a}})\citenamefont {Abbott} \emph
  {et~al.}}]{LIGOScientific:2016aoc}%
  \BibitemOpen
  \bibfield  {author} {\bibinfo {author} {\bibfnamefont {B.~P.}\ \bibnamefont
  {Abbott}} \emph {et~al.} (\bibinfo {collaboration} {LIGO Scientific,
  Virgo}),\ }\href {\doibase 10.1103/PhysRevLett.116.061102} {\bibfield
  {journal} {\bibinfo  {journal} {Phys. Rev. Lett.}\ }\textbf {\bibinfo
  {volume} {116}},\ \bibinfo {pages} {061102} (\bibinfo {year}
  {2016}{\natexlab{a}})},\ \Eprint {http://arxiv.org/abs/1602.03837}
  {arXiv:1602.03837 [gr-qc]} \BibitemShut {NoStop}%
\bibitem [{\citenamefont {Acernese}\ \emph {et~al.}(2015)\citenamefont
  {Acernese} \emph {et~al.}}]{VIRGO:2014yos}%
  \BibitemOpen
  \bibfield  {author} {\bibinfo {author} {\bibfnamefont {F.}~\bibnamefont
  {Acernese}} \emph {et~al.} (\bibinfo {collaboration} {VIRGO}),\ }\href
  {\doibase 10.1088/0264-9381/32/2/024001} {\bibfield  {journal} {\bibinfo
  {journal} {Class. Quant. Grav.}\ }\textbf {\bibinfo {volume} {32}},\ \bibinfo
  {pages} {024001} (\bibinfo {year} {2015})},\ \Eprint
  {http://arxiv.org/abs/1408.3978} {arXiv:1408.3978 [gr-qc]} \BibitemShut
  {NoStop}%
\bibitem [{\citenamefont {Akutsu}\ \emph {et~al.}(2020)\citenamefont {Akutsu}
  \emph {et~al.}}]{KAGRA:2020cvd}%
  \BibitemOpen
  \bibfield  {author} {\bibinfo {author} {\bibfnamefont {T.}~\bibnamefont
  {Akutsu}} \emph {et~al.} (\bibinfo {collaboration} {KAGRA}),\ }\href
  {\doibase 10.1093/ptep/ptaa120} {\  (\bibinfo {year} {2020}),\
  10.1093/ptep/ptaa120},\ \Eprint {http://arxiv.org/abs/2008.02921}
  {arXiv:2008.02921 [gr-qc]} \BibitemShut {NoStop}%
\bibitem [{\citenamefont {Abbott}\ \emph {et~al.}(2019)\citenamefont {Abbott}
  \emph {et~al.}}]{LIGOScientific:2018mvr}%
  \BibitemOpen
  \bibfield  {author} {\bibinfo {author} {\bibfnamefont {B.~P.}\ \bibnamefont
  {Abbott}} \emph {et~al.} (\bibinfo {collaboration} {LIGO Scientific,
  Virgo}),\ }\href {\doibase 10.1103/PhysRevX.9.031040} {\bibfield  {journal}
  {\bibinfo  {journal} {Phys. Rev. X}\ }\textbf {\bibinfo {volume} {9}},\
  \bibinfo {pages} {031040} (\bibinfo {year} {2019})},\ \Eprint
  {http://arxiv.org/abs/1811.12907} {arXiv:1811.12907 [astro-ph.HE]}
  \BibitemShut {NoStop}%
\bibitem [{\citenamefont {Abbott}\ \emph
  {et~al.}(2021{\natexlab{a}})\citenamefont {Abbott} \emph
  {et~al.}}]{LIGOScientific:2020ibl}%
  \BibitemOpen
  \bibfield  {author} {\bibinfo {author} {\bibfnamefont {R.}~\bibnamefont
  {Abbott}} \emph {et~al.} (\bibinfo {collaboration} {LIGO Scientific,
  Virgo}),\ }\href {\doibase 10.1103/PhysRevX.11.021053} {\bibfield  {journal}
  {\bibinfo  {journal} {Phys. Rev. X}\ }\textbf {\bibinfo {volume} {11}},\
  \bibinfo {pages} {021053} (\bibinfo {year} {2021}{\natexlab{a}})},\ \Eprint
  {http://arxiv.org/abs/2010.14527} {arXiv:2010.14527 [gr-qc]} \BibitemShut
  {NoStop}%
\bibitem [{\citenamefont {Abbott}\ \emph
  {et~al.}(2021{\natexlab{b}})\citenamefont {Abbott} \emph
  {et~al.}}]{LIGOScientific:2021djp}%
  \BibitemOpen
  \bibfield  {author} {\bibinfo {author} {\bibfnamefont {R.}~\bibnamefont
  {Abbott}} \emph {et~al.} (\bibinfo {collaboration} {LIGO Scientific, VIRGO,
  KAGRA}),\ }\href@noop {} {\  (\bibinfo {year} {2021}{\natexlab{b}})},\
  \Eprint {http://arxiv.org/abs/2111.03606} {arXiv:2111.03606 [gr-qc]}
  \BibitemShut {NoStop}%
\bibitem [{\citenamefont {Abbott}\ \emph {et~al.}(2017)\citenamefont {Abbott}
  \emph {et~al.}}]{LIGOScientific:2017vwq}%
  \BibitemOpen
  \bibfield  {author} {\bibinfo {author} {\bibfnamefont {B.~P.}\ \bibnamefont
  {Abbott}} \emph {et~al.} (\bibinfo {collaboration} {LIGO Scientific,
  Virgo}),\ }\href {\doibase 10.1103/PhysRevLett.119.161101} {\bibfield
  {journal} {\bibinfo  {journal} {Phys. Rev. Lett.}\ }\textbf {\bibinfo
  {volume} {119}},\ \bibinfo {pages} {161101} (\bibinfo {year} {2017})},\
  \Eprint {http://arxiv.org/abs/1710.05832} {arXiv:1710.05832 [gr-qc]}
  \BibitemShut {NoStop}%
\bibitem [{\citenamefont {Abbott}\ \emph {et~al.}(2020)\citenamefont {Abbott}
  \emph {et~al.}}]{LIGOScientific:2020aai}%
  \BibitemOpen
  \bibfield  {author} {\bibinfo {author} {\bibfnamefont {B.~P.}\ \bibnamefont
  {Abbott}} \emph {et~al.} (\bibinfo {collaboration} {LIGO Scientific,
  Virgo}),\ }\href {\doibase 10.3847/2041-8213/ab75f5} {\bibfield  {journal}
  {\bibinfo  {journal} {Astrophys. J. Lett.}\ }\textbf {\bibinfo {volume}
  {892}},\ \bibinfo {pages} {L3} (\bibinfo {year} {2020})},\ \Eprint
  {http://arxiv.org/abs/2001.01761} {arXiv:2001.01761 [astro-ph.HE]}
  \BibitemShut {NoStop}%
\bibitem [{\citenamefont {Shibata}\ \emph {et~al.}(2017)\citenamefont
  {Shibata}, \citenamefont {Fujibayashi}, \citenamefont {Hotokezaka},
  \citenamefont {Kiuchi}, \citenamefont {Kyutoku}, \citenamefont {Sekiguchi},\
  and\ \citenamefont {Tanaka}}]{Shibata:2017xdx}%
  \BibitemOpen
  \bibfield  {author} {\bibinfo {author} {\bibfnamefont {M.}~\bibnamefont
  {Shibata}}, \bibinfo {author} {\bibfnamefont {S.}~\bibnamefont
  {Fujibayashi}}, \bibinfo {author} {\bibfnamefont {K.}~\bibnamefont
  {Hotokezaka}}, \bibinfo {author} {\bibfnamefont {K.}~\bibnamefont {Kiuchi}},
  \bibinfo {author} {\bibfnamefont {K.}~\bibnamefont {Kyutoku}}, \bibinfo
  {author} {\bibfnamefont {Y.}~\bibnamefont {Sekiguchi}}, \ and\ \bibinfo
  {author} {\bibfnamefont {M.}~\bibnamefont {Tanaka}},\ }\href {\doibase
  10.1103/PhysRevD.96.123012} {\bibfield  {journal} {\bibinfo  {journal} {Phys.
  Rev. D}\ }\textbf {\bibinfo {volume} {96}},\ \bibinfo {pages} {123012}
  (\bibinfo {year} {2017})},\ \Eprint {http://arxiv.org/abs/1710.07579}
  {arXiv:1710.07579 [astro-ph.HE]} \BibitemShut {NoStop}%
\bibitem [{\citenamefont {Hannam}\ \emph {et~al.}(2014)\citenamefont {Hannam},
  \citenamefont {Schmidt}, \citenamefont {Boh\'e}, \citenamefont {Haegel},
  \citenamefont {Husa}, \citenamefont {Ohme}, \citenamefont {Pratten},\ and\
  \citenamefont {P\"urrer}}]{Hannam:2013oca}%
  \BibitemOpen
  \bibfield  {author} {\bibinfo {author} {\bibfnamefont {M.}~\bibnamefont
  {Hannam}}, \bibinfo {author} {\bibfnamefont {P.}~\bibnamefont {Schmidt}},
  \bibinfo {author} {\bibfnamefont {A.}~\bibnamefont {Boh\'e}}, \bibinfo
  {author} {\bibfnamefont {L.}~\bibnamefont {Haegel}}, \bibinfo {author}
  {\bibfnamefont {S.}~\bibnamefont {Husa}}, \bibinfo {author} {\bibfnamefont
  {F.}~\bibnamefont {Ohme}}, \bibinfo {author} {\bibfnamefont {G.}~\bibnamefont
  {Pratten}}, \ and\ \bibinfo {author} {\bibfnamefont {M.}~\bibnamefont
  {P\"urrer}},\ }\href {\doibase 10.1103/PhysRevLett.113.151101} {\bibfield
  {journal} {\bibinfo  {journal} {Phys. Rev. Lett.}\ }\textbf {\bibinfo
  {volume} {113}},\ \bibinfo {pages} {151101} (\bibinfo {year} {2014})},\
  \Eprint {http://arxiv.org/abs/1308.3271} {arXiv:1308.3271 [gr-qc]}
  \BibitemShut {NoStop}%
\bibitem [{\citenamefont {Boh\'e}\ \emph {et~al.}(2017)\citenamefont {Boh\'e}
  \emph {et~al.}}]{Bohe:2016gbl}%
  \BibitemOpen
  \bibfield  {author} {\bibinfo {author} {\bibfnamefont {A.}~\bibnamefont
  {Boh\'e}} \emph {et~al.},\ }\href {\doibase 10.1103/PhysRevD.95.044028}
  {\bibfield  {journal} {\bibinfo  {journal} {Phys. Rev. D}\ }\textbf {\bibinfo
  {volume} {95}},\ \bibinfo {pages} {044028} (\bibinfo {year} {2017})},\
  \Eprint {http://arxiv.org/abs/1611.03703} {arXiv:1611.03703 [gr-qc]}
  \BibitemShut {NoStop}%
\bibitem [{\citenamefont {Khan}\ \emph {et~al.}(2016)\citenamefont {Khan},
  \citenamefont {Husa}, \citenamefont {Hannam}, \citenamefont {Ohme},
  \citenamefont {P\"urrer}, \citenamefont {Jim\'enez~Forteza},\ and\
  \citenamefont {Boh\'e}}]{Khan:2015jqa}%
  \BibitemOpen
  \bibfield  {author} {\bibinfo {author} {\bibfnamefont {S.}~\bibnamefont
  {Khan}}, \bibinfo {author} {\bibfnamefont {S.}~\bibnamefont {Husa}}, \bibinfo
  {author} {\bibfnamefont {M.}~\bibnamefont {Hannam}}, \bibinfo {author}
  {\bibfnamefont {F.}~\bibnamefont {Ohme}}, \bibinfo {author} {\bibfnamefont
  {M.}~\bibnamefont {P\"urrer}}, \bibinfo {author} {\bibfnamefont
  {X.}~\bibnamefont {Jim\'enez~Forteza}}, \ and\ \bibinfo {author}
  {\bibfnamefont {A.}~\bibnamefont {Boh\'e}},\ }\href {\doibase
  10.1103/PhysRevD.93.044007} {\bibfield  {journal} {\bibinfo  {journal} {Phys.
  Rev. D}\ }\textbf {\bibinfo {volume} {93}},\ \bibinfo {pages} {044007}
  (\bibinfo {year} {2016})},\ \Eprint {http://arxiv.org/abs/1508.07253}
  {arXiv:1508.07253 [gr-qc]} \BibitemShut {NoStop}%
\bibitem [{\citenamefont {Blackman}\ \emph {et~al.}(2017)\citenamefont
  {Blackman}, \citenamefont {Field}, \citenamefont {Scheel}, \citenamefont
  {Galley}, \citenamefont {Ott}, \citenamefont {Boyle}, \citenamefont {Kidder},
  \citenamefont {Pfeiffer},\ and\ \citenamefont
  {Szil\'agyi}}]{Blackman:2017pcm}%
  \BibitemOpen
  \bibfield  {author} {\bibinfo {author} {\bibfnamefont {J.}~\bibnamefont
  {Blackman}}, \bibinfo {author} {\bibfnamefont {S.~E.}\ \bibnamefont {Field}},
  \bibinfo {author} {\bibfnamefont {M.~A.}\ \bibnamefont {Scheel}}, \bibinfo
  {author} {\bibfnamefont {C.~R.}\ \bibnamefont {Galley}}, \bibinfo {author}
  {\bibfnamefont {C.~D.}\ \bibnamefont {Ott}}, \bibinfo {author} {\bibfnamefont
  {M.}~\bibnamefont {Boyle}}, \bibinfo {author} {\bibfnamefont {L.~E.}\
  \bibnamefont {Kidder}}, \bibinfo {author} {\bibfnamefont {H.~P.}\
  \bibnamefont {Pfeiffer}}, \ and\ \bibinfo {author} {\bibfnamefont
  {B.}~\bibnamefont {Szil\'agyi}},\ }\href {\doibase
  10.1103/PhysRevD.96.024058} {\bibfield  {journal} {\bibinfo  {journal} {Phys.
  Rev. D}\ }\textbf {\bibinfo {volume} {96}},\ \bibinfo {pages} {024058}
  (\bibinfo {year} {2017})},\ \Eprint {http://arxiv.org/abs/1705.07089}
  {arXiv:1705.07089 [gr-qc]} \BibitemShut {NoStop}%
\bibitem [{\citenamefont {Husa}\ \emph {et~al.}(2016)\citenamefont {Husa},
  \citenamefont {Khan}, \citenamefont {Hannam}, \citenamefont {P\"urrer},
  \citenamefont {Ohme}, \citenamefont {Jim\'enez~Forteza},\ and\ \citenamefont
  {Boh\'e}}]{Husa:2015iqa}%
  \BibitemOpen
  \bibfield  {author} {\bibinfo {author} {\bibfnamefont {S.}~\bibnamefont
  {Husa}}, \bibinfo {author} {\bibfnamefont {S.}~\bibnamefont {Khan}}, \bibinfo
  {author} {\bibfnamefont {M.}~\bibnamefont {Hannam}}, \bibinfo {author}
  {\bibfnamefont {M.}~\bibnamefont {P\"urrer}}, \bibinfo {author}
  {\bibfnamefont {F.}~\bibnamefont {Ohme}}, \bibinfo {author} {\bibfnamefont
  {X.}~\bibnamefont {Jim\'enez~Forteza}}, \ and\ \bibinfo {author}
  {\bibfnamefont {A.}~\bibnamefont {Boh\'e}},\ }\href {\doibase
  10.1103/PhysRevD.93.044006} {\bibfield  {journal} {\bibinfo  {journal} {Phys.
  Rev. D}\ }\textbf {\bibinfo {volume} {93}},\ \bibinfo {pages} {044006}
  (\bibinfo {year} {2016})},\ \Eprint {http://arxiv.org/abs/1508.07250}
  {arXiv:1508.07250 [gr-qc]} \BibitemShut {NoStop}%
\bibitem [{\citenamefont {Abbott}\ \emph
  {et~al.}(2016{\natexlab{b}})\citenamefont {Abbott} \emph
  {et~al.}}]{LIGOScientific:2016fbo}%
  \BibitemOpen
  \bibfield  {author} {\bibinfo {author} {\bibfnamefont {B.~P.}\ \bibnamefont
  {Abbott}} \emph {et~al.} (\bibinfo {collaboration} {LIGO Scientific,
  Virgo}),\ }\href {\doibase 10.1103/PhysRevD.93.122004} {\bibfield  {journal}
  {\bibinfo  {journal} {Phys. Rev. D}\ }\textbf {\bibinfo {volume} {93}},\
  \bibinfo {pages} {122004} (\bibinfo {year} {2016}{\natexlab{b}})},\ \bibinfo
  {note} {[Addendum: Phys.Rev.D 94, 069903 (2016)]},\ \Eprint
  {http://arxiv.org/abs/1602.03843} {arXiv:1602.03843 [gr-qc]} \BibitemShut
  {NoStop}%
\bibitem [{\citenamefont {Lange}\ \emph {et~al.}(2017)\citenamefont {Lange}
  \emph {et~al.}}]{Lange:2017wki}%
  \BibitemOpen
  \bibfield  {author} {\bibinfo {author} {\bibfnamefont {J.}~\bibnamefont
  {Lange}} \emph {et~al.},\ }\href {\doibase 10.1103/PhysRevD.96.104041}
  {\bibfield  {journal} {\bibinfo  {journal} {Phys. Rev. D}\ }\textbf {\bibinfo
  {volume} {96}},\ \bibinfo {pages} {104041} (\bibinfo {year} {2017})},\
  \Eprint {http://arxiv.org/abs/1705.09833} {arXiv:1705.09833 [gr-qc]}
  \BibitemShut {NoStop}%
\bibitem [{\citenamefont {Schmidt}\ \emph {et~al.}(2017)\citenamefont
  {Schmidt}, \citenamefont {Harry},\ and\ \citenamefont
  {Pfeiffer}}]{Schmidt:2017btt}%
  \BibitemOpen
  \bibfield  {author} {\bibinfo {author} {\bibfnamefont {P.}~\bibnamefont
  {Schmidt}}, \bibinfo {author} {\bibfnamefont {I.~W.}\ \bibnamefont {Harry}},
  \ and\ \bibinfo {author} {\bibfnamefont {H.~P.}\ \bibnamefont {Pfeiffer}},\
  }\href@noop {} {\  (\bibinfo {year} {2017})},\ \Eprint
  {http://arxiv.org/abs/1703.01076} {arXiv:1703.01076 [gr-qc]} \BibitemShut
  {NoStop}%
\bibitem [{\citenamefont {Loffler}\ \emph {et~al.}(2012)\citenamefont {Loffler}
  \emph {et~al.}}]{Loffler:2011ay}%
  \BibitemOpen
  \bibfield  {author} {\bibinfo {author} {\bibfnamefont {F.}~\bibnamefont
  {Loffler}} \emph {et~al.},\ }\href {\doibase 10.1088/0264-9381/29/11/115001}
  {\bibfield  {journal} {\bibinfo  {journal} {Class. Quant. Grav.}\ }\textbf
  {\bibinfo {volume} {29}},\ \bibinfo {pages} {115001} (\bibinfo {year}
  {2012})},\ \Eprint {http://arxiv.org/abs/1111.3344} {arXiv:1111.3344 [gr-qc]}
  \BibitemShut {NoStop}%
\bibitem [{\citenamefont {Herrmann}\ \emph {et~al.}(2006)\citenamefont
  {Herrmann}, \citenamefont {Hinder}, \citenamefont {Shoemaker},\ and\
  \citenamefont {Laguna}}]{Herrmann:2006ks}%
  \BibitemOpen
  \bibfield  {author} {\bibinfo {author} {\bibfnamefont {F.}~\bibnamefont
  {Herrmann}}, \bibinfo {author} {\bibfnamefont {I.}~\bibnamefont {Hinder}},
  \bibinfo {author} {\bibfnamefont {D.}~\bibnamefont {Shoemaker}}, \ and\
  \bibinfo {author} {\bibfnamefont {P.}~\bibnamefont {Laguna}},\ }in\
  \href@noop {} {\emph {\bibinfo {booktitle} {{New Frontiers in Numerical
  Relativity (NFNR 2006)}}}}\ (\bibinfo {year} {2006})\ \Eprint
  {http://arxiv.org/abs/gr-qc/0601026} {arXiv:gr-qc/0601026} \BibitemShut
  {NoStop}%
\bibitem [{\citenamefont {Jani}\ \emph {et~al.}(2016)\citenamefont {Jani},
  \citenamefont {Healy}, \citenamefont {Clark}, \citenamefont {London},
  \citenamefont {Laguna},\ and\ \citenamefont {Shoemaker}}]{Jani:2016wkt}%
  \BibitemOpen
  \bibfield  {author} {\bibinfo {author} {\bibfnamefont {K.}~\bibnamefont
  {Jani}}, \bibinfo {author} {\bibfnamefont {J.}~\bibnamefont {Healy}},
  \bibinfo {author} {\bibfnamefont {J.~A.}\ \bibnamefont {Clark}}, \bibinfo
  {author} {\bibfnamefont {L.}~\bibnamefont {London}}, \bibinfo {author}
  {\bibfnamefont {P.}~\bibnamefont {Laguna}}, \ and\ \bibinfo {author}
  {\bibfnamefont {D.}~\bibnamefont {Shoemaker}},\ }\href {\doibase
  10.1088/0264-9381/33/20/204001} {\bibfield  {journal} {\bibinfo  {journal}
  {Class. Quant. Grav.}\ }\textbf {\bibinfo {volume} {33}},\ \bibinfo {pages}
  {204001} (\bibinfo {year} {2016})},\ \Eprint
  {http://arxiv.org/abs/1605.03204} {arXiv:1605.03204 [gr-qc]} \BibitemShut
  {NoStop}%
\bibitem [{\citenamefont {Vaishnav}\ \emph {et~al.}(2007)\citenamefont
  {Vaishnav}, \citenamefont {Hinder}, \citenamefont {Herrmann},\ and\
  \citenamefont {Shoemaker}}]{Vaishnav:2007nm}%
  \BibitemOpen
  \bibfield  {author} {\bibinfo {author} {\bibfnamefont {B.}~\bibnamefont
  {Vaishnav}}, \bibinfo {author} {\bibfnamefont {I.}~\bibnamefont {Hinder}},
  \bibinfo {author} {\bibfnamefont {F.}~\bibnamefont {Herrmann}}, \ and\
  \bibinfo {author} {\bibfnamefont {D.}~\bibnamefont {Shoemaker}},\ }\href
  {\doibase 10.1103/PhysRevD.76.084020} {\bibfield  {journal} {\bibinfo
  {journal} {Phys. Rev. D}\ }\textbf {\bibinfo {volume} {76}},\ \bibinfo
  {pages} {084020} (\bibinfo {year} {2007})},\ \Eprint
  {http://arxiv.org/abs/0705.3829} {arXiv:0705.3829 [gr-qc]} \BibitemShut
  {NoStop}%
\bibitem [{\citenamefont {Healy}\ \emph {et~al.}(2009)\citenamefont {Healy},
  \citenamefont {Levin},\ and\ \citenamefont {Shoemaker}}]{Healy:2009zm}%
  \BibitemOpen
  \bibfield  {author} {\bibinfo {author} {\bibfnamefont {J.}~\bibnamefont
  {Healy}}, \bibinfo {author} {\bibfnamefont {J.}~\bibnamefont {Levin}}, \ and\
  \bibinfo {author} {\bibfnamefont {D.}~\bibnamefont {Shoemaker}},\ }\href
  {\doibase 10.1103/PhysRevLett.103.131101} {\bibfield  {journal} {\bibinfo
  {journal} {Phys. Rev. Lett.}\ }\textbf {\bibinfo {volume} {103}},\ \bibinfo
  {pages} {131101} (\bibinfo {year} {2009})},\ \Eprint
  {http://arxiv.org/abs/0907.0671} {arXiv:0907.0671 [gr-qc]} \BibitemShut
  {NoStop}%
\bibitem [{\citenamefont {Pekowsky}\ \emph {et~al.}(2013)\citenamefont
  {Pekowsky}, \citenamefont {O'Shaughnessy}, \citenamefont {Healy},\ and\
  \citenamefont {Shoemaker}}]{Pekowsky:2013ska}%
  \BibitemOpen
  \bibfield  {author} {\bibinfo {author} {\bibfnamefont {L.}~\bibnamefont
  {Pekowsky}}, \bibinfo {author} {\bibfnamefont {R.}~\bibnamefont
  {O'Shaughnessy}}, \bibinfo {author} {\bibfnamefont {J.}~\bibnamefont
  {Healy}}, \ and\ \bibinfo {author} {\bibfnamefont {D.}~\bibnamefont
  {Shoemaker}},\ }\href {\doibase 10.1103/PhysRevD.88.024040} {\bibfield
  {journal} {\bibinfo  {journal} {Phys. Rev. D}\ }\textbf {\bibinfo {volume}
  {88}},\ \bibinfo {pages} {024040} (\bibinfo {year} {2013})},\ \Eprint
  {http://arxiv.org/abs/1304.3176} {arXiv:1304.3176 [gr-qc]} \BibitemShut
  {NoStop}%
\end{thebibliography}%
\end{document}